\documentclass[aps,twocolumn,amssymb,nobibnotes,nofootinbib,longbibliography,superscriptaddress,10pt]{revtex4-1}

\usepackage{titlesec}
\usepackage{bbold}

\usepackage{graphicx,epsf}
\usepackage{amsmath,amsfonts,amssymb,amsbsy,mathrsfs}

\usepackage{caption}
\usepackage{subcaption}
\usepackage{physics}

\usepackage{color,xcolor}

\usepackage{array,multirow}
\usepackage{stackrel}

\usepackage{ifdraft}
\usepackage[obeyFinal]{easy-todo}

\usepackage{placeins}
\usepackage{slashed,shuffle} 
\usepackage{adjustbox}

\usepackage{comment}
\usepackage[normalem]{ulem}

\usepackage{slashed}
\usepackage{hyperref}
\hypersetup{
    colorlinks=true,
    allcolors={blue!60!black}
}

\usepackage{cleveref}
\usepackage{pifont}

\newcolumntype{C}[1]{>{\centering\arraybackslash}m{#1}}

\titleformat{\section}{\centering\normalsize\normalfont\bf}{\thesection}{1em}{}
\makeatletter\DeclareRobustCommand*{\bfseries}{\not@math@alphabet\bfseries\mathbf\fontseries\bfdefault\selectfont\boldmath}\makeatother

\def\RU{Institute for Theoretical Physics, University of Regensburg, 93040 Regensburg, Germany}

\begin{document}

\title{Phase-space integrals through Mellin-Barnes representation}

\author{Taushif Ahmed}
\email[Electronic address: ]{taushif.ahmed@ur.de}
\affiliation{\RU}

\author{Syed Mehedi Hasan}
\email[Electronic address: ]{syed-mehedi.hasan@ur.de}
\affiliation{\RU}

\author{Andreas Rapakoulias}
\email[Electronic address: ]{andreas.rapakoulias@ur.de}
\affiliation{\RU}


\begin{abstract}
This letter introduces a novel analytical approach to calculating phase-space integrals, crucial for precision in particle physics. We develop a method to compute angular components using multifold Mellin-Barnes integrals, yielding results in terms of Goncharov polylogarithms for integrals involving three denominators. Our results include expressions for massless momenta up to $\mathcal{O}(\epsilon^2)$ and for one massive momentum up to $\mathcal{O}(\epsilon)$. Additionally, we derive recursion relations that reduce integrals with higher powers of denominators to simpler ones. We detail how to combine the angular part with the radial one which requires a careful handling of singularities.

\end{abstract}

\maketitle



%

Phase-space (PS) integrals are essential for achieving precise theoretical predictions in perturbative quantum field theory owing to their appearance through real-emission Feynman diagrams. Calculating an observable beyond the leading perturbative order requires consideration of both virtual and real emission diagrams. Over the past few decades, a plethora of work has been performed in handling Feynman integrals analytically arising in virtual diagrams. In contrast, the analytical computation of PS integrals has not received comparable attention, despite being a crucial component. In this letter, we present a method based on the Mellin-Barnes (MB) representation to perform an efficient analytical computation of PS integrals within the framework of dimensional regularization.

Phase-space integrals can be decomposed into radial and angular parts in an appropriately chosen reference frame. In $d=4-2\epsilon$ space-time dimensions, the latter with $n$ denominators can be expressed as 
\begin{align}
\label{eq:ang_int}
\Omega_{j_{1}, \ldots, j_{n}} \equiv \int \mathrm{d} \Omega_{d-1}(q) \frac{1}{\left(p_{1} \cdot q\right)^{j_{1}} \ldots\left(p_{n} \cdot q\right)^{j_{n}}}\,,
\end{align}
where $\{p_i^\mu\}$ are a set of fixed vectors and $\mathrm{d} \Omega_{d-1}(q)$ is rotationally invariant angular measure for the massless vector $q^\mu$. The $\{j_i\}$ are integers whose values depend on the scattering process and the perturbative order of interest. For instance, for the double real emission in semi-inclusive deep inelastic scattering at next-to-next-to-leading order, we require a maximum of two denominators, i.e., $n=2$. 
When $n=1$ and the momentum $p_1$ is massless $(p_1^2=0)$,  the integral is straightforward to express in terms of the total volume element. For massive momentum ($p_1^2\neq 0$), the angular integral can be solved in terms of the $_2F_1$ hypergeometric function~\cite{vanNeerven:1985xr,Somogyi:2011ir}. 
For the case of two denominators with massless momenta ($n=2$ and $p_1^2=p_2^2=0$), the angular integral can again be solved using the $_2F_1$ hypergeometric function~\cite{vanNeerven:1985xr,Somogyi:2011ir}. However, when one or both momenta are massive, the result requires Appell and Fox's H-functions, respectively~\cite{Somogyi:2011ir}. 
Integrals with three or more denominators can be expressed in terms of the H-function of several variables as an all-order expression in 
$\epsilon$. However, the evaluation of this function in powers of $\epsilon$ is a highly non-trivial task. In this letter, we demonstrate a method based on the MB representation to solve these integrals as a Laurent series in the dimensional regulator. Using this method, we solve the angular integrals of the three denominators and present their results up to ${\cal O}(\epsilon^2)$ for massless case and to ${\cal O}(\epsilon)$ for massive cases for the first time. Since the method is algorithmic, it can be used to solve the angular integrals with more number of denominators.  Recently, small-mass asymptotic behaviours of angular integrals of three- and four-denominators have become available~\cite{Smirnov:2024pbj}.

\section{Mellin-Barnes representation}
\label{sec:MB-rep}

The angular integral~\eqref{eq:ang_int} exhibits the following Mellin-Barnes (MB) representation:
\begin{align}
\label{eq:MB}
\Omega_{j_{1}, \ldots, j_{n}}\left(\left\{v_{k l}\right\}, \epsilon\right)=&  \frac{2^{2-j-2 \epsilon} \pi^{1-\epsilon}}{\prod_{k=1}^{n} \Gamma\left(j_{k}\right) \Gamma(2-j-2 \epsilon)} \nonumber\\
 \times \int_{-\mathrm{i} \infty}^{+\mathrm{i} \infty}&\bigg[\prod_{k=1}^{n} \prod_{l=k}^{n} \frac{\mathrm{d} z_{k l}}{2 \pi \mathrm{i}} \Gamma\left(-z_{k l}\right)\left(v_{k l}\right)^{z_{k l}}\bigg]\nonumber\\
\times \big[\prod_{k=1}^{n} \Gamma&\left(j_{k}+z_{k}\right)\big] \Gamma(1-j-\epsilon-z)\,.
\end{align}
The scalar products among $p_i^\mu$'s are defined through
\begin{align}
v_{k l} \equiv \begin{cases}\frac{p_{k} \cdot p_{l}}{2}& ~{\rm for}~ k \neq l \\ \frac{p_{k}^{2}}{4} & ~{\rm for}~ k=l\end{cases}.
\end{align}
The MB integrals are over the variables $z_{kl}$ from which we further define 
\begin{align}
z=\sum_{k=1}^{n} \sum_{l=k}^{n} z_{k l} \quad \text { and } \quad z_{k}=\sum_{l=1}^{k} z_{l k}+\sum_{l=k}^{n} z_{k l}.
\end{align}
In words, $z$ represents the total sum of all MB variables. For each index $k$, $z_k$ denotes the sum of all variables involving $k$ as one of their indices, with $z_{kk}$ being counted twice. Specifically, this can be expressed as $z_k = z_{1k} + \ldots+z_{k-1 k} + 2z_{kk} +z_{k k+1 }+ \ldots +z_{kn}$. The variable $j$ is sum of all $j_k$'s, i.e., $j=\sum_{k=1}^n j_k$. 
In deriving the MB representation, it is implicitly assumed that the indices, $j_k$'s, are positive integers. In case of negative integers, the result can be obtained through analytic continuation. Moreover, in \eqref{eq:MB}, we also assume $v_{kl}\neq 0$. If some of $v_{kl}$ are zero, which could happen for massless momentum $p_i$ for which $v_{ii}=0$, then the corresponding integral over $z_{kl}$ should be omitted and $z_{kl}$ should be put to zero in the rest of the expression.

Except in very limited cases, these MB integrals can not be evaluated in a closed form. Since our goal is to solve the integrals in a Laurent series expansion of $\epsilon$ rather than in terms of an all-order expression, we follow the following strategy. We expand the MB integral around $\epsilon \rightarrow 0$ and then evaluate the coefficients of this expansion, which generally also contain multifold MB integrals. To perform such an expansion safely, it is essential to ensure that, as the integrand approaches the limit $\epsilon \rightarrow 0$, the 
integration contour does not cross any of the poles. This occurs when the real parts of the arguments of the gamma functions in the integrand are not all positive. In such scenarios, an analytic continuation to a region around $\epsilon=0$ becomes inevitable. 
To accomplish this, we must deform the contour~\cite{Smirnov:1999gc} and account for the corresponding residues when the contour crosses any of the poles during this deformation. In the MB representation, the originally chosen integration contours are usually parallel to the imaginary axis, so that all the poles originating from the gamma function of the type $ \Gamma(a+z)$ are on the left and all the poles of gamma functions of the type $ \Gamma(a-z)$ are on the right. Therefore, we have to effectively shift the contour along the real axis and pick up the contributions of the residues when the contour crosses the poles. 
Alternatively, following the method in~\cite{Tausk:1999vh,Czakon:2005rk}, we can maintain a fixed integration contour and track the movement of the gamma function poles as $\epsilon$ varies. Each time a pole crosses an integration contour, its residue is added to the right-hand side of \eqref{eq:MB} for the corresponding variable. The residue terms themselves may contain poles in the remaining variables, which might also cross their respective contours as $\epsilon$ changes. These instances must be handled similarly. Ultimately, this process yields a collection of residue terms in addition to the original multifold MB integral. The final integrand can then be safely expanded, as the contours remain clear of poles as $\epsilon \rightarrow 0$.
This process effectively decomposes the original MB integral into several MB integrals. 
After the $\epsilon$ expansion of each integral is complete, we collect all the terms of all the integrals that are of the same order in $\epsilon$ and obtain a total $\epsilon$ expansion for the original MB integral. The primary task then is to calculate the resulting MB integrals that appear in the coefficients of the expansion. In the following section~\ref{sec:ana-int}, we outline the method for evaluating these MB integrals and represent the results in terms of multiple polylogarithms or iterated integrals. However, when dealing with multifold MB integrals, applying this general approach to fully solve them becomes highly challenging. We illustrate these complexities in section~\ref{sec:3den-massless} by addressing the case of three denominators.

\section{Analytic integration from MB representation}
\label{sec:ana-int}

To calculate the MB integrals that appear in the coefficients of the $\epsilon$ expansion, we observe that the integrals are balanced. We say that a MB integral, 
\begin{equation}
\int_{-i\infty}^{+i\infty}\frac{dz_{j}}{2\pi i}\prod^{n_{+}}_{k=1}\Gamma(a_{k}+z_{j})^{\alpha_{k}}\prod^{n_{-}}_{l=1}\Gamma(a_{l}-z_{j})^{\beta_{l}}, \hspace{10pt} \alpha_{k},\beta_{l} \in \mathbb{Z}  , 
\end{equation}
is balanced in the variable $z_{j}$ if the condition $\sum^{n_{+}}_{k=1} \alpha_{k}=\sum_{l=1}^{n_{-}} \beta_{l}$
is fulfilled. 
A multi-dimensional MB integral is considered balanced if each integration variable satisfies this condition individually. For a balanced MB integral, the integrand can be expressed as a product of beta functions. This is made possible by ensuring that the gamma functions have positive real parts, which is typically achieved after performing analytic continuation and expanding around $\epsilon=0$. In such cases, all gamma functions can be combined into beta functions, ensuring that none of the remaining gamma functions contain integration variables. We then replace the beta functions with their integral representations, selecting the appropriate one based on the specific context:
\begin{equation}
    B(a,b)=
\begin{cases}
\int_{0}^{1} dx x^{a-1}(1-x)^{b-1},\\
\int_{0}^{\infty} dx x^{a-1}(1+x)^{-a-b},
\end{cases}
\end{equation}
where $\mathcal{R}(a)>0$ and $\mathcal{R}(b)>0$. Since the resulting integral is convergent, we can safely interchange the order of the MB integration with the real integrations arising from the beta functions. This allows us to recast the integral into 
\begin{equation}
\label{eq:real-mb}
\int^{\kappa}_{0} \Bigg( \prod^{K}_{i=1} dx_{i} \Bigg) R_{0}(\mathbf{x},\mathbf{u})\int^{+i\infty}_{-i\infty}\prod^{L}_{l=1} \frac{d\tilde z_{l}}{2\pi i}\Bigg[ R_{l}(\mathbf{x},\mathbf{u})\Bigg]^{\tilde z_{l}},
\end{equation}
where $\mathbf{x}=(x_{1},...,x_{K})$ are $K$ real integration variables originating from the beta functions, $\mathbf{u}=(u_{1},...,u_{N})$ are the parameters other than the integration variables present in the original MB integral and the $R_{0},...,R_{l}$ are ratios of products of $x_{i}$, $(1\pm x_{i})$ and $u_{j}$. The upper limit, $\kappa$, is either $1$ or $\infty$. The MB integral in \eqref{eq:real-mb} can now be readily evaluated, reducing the problem to solving a few integrals over real parameters, $\mathbf{x}$. The residual integrand typically consists of rational functions and logarithms of polynomials in the integration variables. Ideally, if these integrals can be performed iteratively, the result can be expressed in terms of multiple polylogarithms (MPL) or iterated integrals. However, if the integrand involves more complex structures, it may not be possible to represent the result in terms of MPLs. In the case of multifold MB integrals, this process involves significant difficulties. In the next section, \ref{sec:3den-massless}, we illustrate how, by following this approach, we successfully solve the case of three propagator angular phase-space integrals in terms of MPLs.

\section{Massless integrals with three denominators}
\label{sec:3den-massless}

We consider the case of three denominators with massless $p_i$'s for which the MB representation \eqref{eq:MB} becomes
\begin{align}
\label{eq:MB-3denom}
\Omega_{j_1,j_2,j_3}&(v_{12},v_{13},v_{23},\epsilon)=\frac{2^{2-j-2\epsilon}\pi^{1-\epsilon}}{\Gamma(j_1)\Gamma(j_2)\Gamma(j_3)\Gamma(2-j-2\epsilon)}& \nonumber\\
&\times \int^{+i\infty}_{-i\infty} \frac{dz_{12}dz_{13}dz_{23}}{(2\pi i)^{3}} \Gamma(-z_{12}) \Gamma(-z_{13}) \Gamma(-z_{23}) \nonumber\\
&\times\Gamma(j_1+z_{12}+z_{13}) \Gamma(j_2+z_{12}+z_{23}) \nonumber\\
&\times\Gamma(j_3+z_{13}+z_{23}) \Gamma(1-j-\epsilon-z) \nonumber\\
&\times(v_{12})^{z_{12}}(v_{13})^{z_{13}} (v_{23})^{z_{23}}
\end{align}
with $z=z_{12}+z_{13}+z_{23}$ and $j=j_1+j_2+j_3$.
For convenience, we include a normalization factor and calculate
\begin{align}
\label{eq:MB-3denom-norm}
I^{(0)}_{j_1,j_2,j_3}&=2^{-1+2\epsilon}\pi^{\epsilon}\frac{\Gamma(1-2\epsilon)}{\Gamma(1-\epsilon)}\Omega_{j_1,j_2,j_3}(v_{12},v_{13},v_{23}).
\end{align}
The superscript $(0)$ indicates that the integral is massless. We start by setting $j_i = 1$, and in section~\ref{sec:recursion}, we extend the discussion to the case of general $j_i$'s. As outlined in section~\ref{sec:MB-rep}, when the integration contours cross the poles as $\epsilon \rightarrow 0$, analytic continuation of the integrand becomes necessary. We then expand the integrand, retaining terms up to $\mathcal{O}(\epsilon^2)$. This procedure yields four distinct types of MB integrals at $\mathcal{O}(\epsilon)$ and eight at $\mathcal{O}(\epsilon^2)$. Of the latter, only four are genuinely new types compared to the former. These integrals share the same arguments but are multiplied by additional polygamma functions. Up to this order, we do not encounter any square roots over the integration variables.
Following the methods outlined in section~\ref{sec:ana-int}, we rewrite the resulting MB integrals in the form presented in \eqref{eq:real-mb}.
By utilizing the identity
\begin{equation}
    \int^{+i\infty+z_{0}}_{-i\infty+z_{0}} \frac{dz}{2\pi i} A^{z} = \delta(1-A), \hspace{10pt} A>0,
\end{equation}
we evaluate the remaining MB integral, ultimately reducing it to a integration over real variables.
\begin{equation}
\int^{1}_{0} \Bigg( \prod^{K}_{i=1} dx_{i} \Bigg) R_{0}(\mathbf{x},\mathbf{u}) \prod^{L}_{l=1} \delta \Bigg[1 - R_{l}(\mathbf{x},\mathbf{u})\Bigg].
\end{equation}
Since the integrand consists of rational functions and logarithms of polynomials, we expect iterated integrals to yield Goncharov polylogarithms (GPLs)\cite{Vollinga:2004sn}. However, in practice, this process is far from straightforward due to the complexity of the integrands. Once a GPL appears in our expressions, we systematically shift the integration variables to the right-most argument of the GPL. For simpler cases involving linear arguments, we use the fibration basis feature implemented in {\tt PolyLogTools}\cite{Duhr:2019tlz}.
However, in many cases, the integration variables appear in the weights of the GPLs as non-linear rational functions, making their manipulation significantly more challenging. Our approach involves first reducing the weight by differentiation, then shifting the variable to the right, and finally reconstructing the original expression through iterative integration~\cite{Gehrmann:2001jv,DelDuca:2009ac,DelDuca:2010zg,Bolzoni:2009ye}. Using in-house algorithms, we automate this process and obtain the final result in terms of GPLs, which we then verify numerically. We provide the analytical result of $I^{(0)}_{1,1,1}$ to ${\cal O}(\epsilon^2)$ in the ancillary file {\tt <3prop0mass.m>}\,.

\section{Massive integrals with three denominators}
The MB representation of the angular integral with one of the $\{p_i\}$'s massive, say $p_1$, and $j_i=1$ takes the form
\begin{align}
\label{eq:MB-3denom-massive}
\Omega_{1,1,1}&(v_{11},v_{12},v_{13},v_{23},\epsilon)=\frac{2^{-1-2\epsilon}\pi^{1-\epsilon}}{\Gamma(-1-2\epsilon)}& \nonumber\\
&\times \int^{+i\infty}_{-i\infty} \frac{dz_{11}dz_{12}dz_{13} dz_{23}}{(2\pi i)^{4}} \Gamma(-z_{11})\Gamma(-z_{12}) \Gamma(-z_{13})\nonumber\\
&\times \Gamma(-z_{23}) 
\Gamma(1+2 z_{11}+z_{12}+z_{13}) \Gamma(1+z_{12}+z_{23}) \nonumber\\
&\times\Gamma(1+z_{13}+z_{23}) \Gamma(-2-\epsilon-z_{11}-z_{12}-z_{13}-z_{23}) \nonumber\\
&\times(v_{11})^{z_{11}}(v_{12})^{z_{12}}(v_{13})^{z_{13}} (v_{23})^{z_{23}}.
\end{align}
Similar to the massless case, we include the same normalisation factor and define
\begin{align}
\label{eq:MB-3denom-norm}
I^{(1)}_{1,1,1}&=2^{-1+2\epsilon}\pi^{\epsilon}\frac{\Gamma(1-2\epsilon)}{\Gamma(1-\epsilon)}\Omega_{1,1,1}(v_{11},v_{12},v_{13},v_{23}),
\end{align}
where the superscript $(1)$ indicates that one of the denominators involves massive momentum. After performing an analytic continuation and expanding to ${\cal O}(\epsilon)$, we obtain eleven distinct MB integrals. Ten of these are evaluated using a method similar to that used in the massless case, with no square roots over the integration variables. However, one MB integral involves a square root over a quadratic function of the integration variable. To address this, we linearize the square root by applying a variable transformation, such that the arguments and coefficients of the resulting GPLs, as well as the Jacobian of the transformation, become rational functions of the new variable. For example, any general quadratic function of $x$, such as $\sqrt{ax^2 +bx +c}$, can be rationalized using the transformation:
\begin{equation}
    x\rightarrow \frac{b+2 c \eta }{c \eta ^2-a}.
\end{equation}
This transformation satisfies the necessary conditions mentioned above. It is important to note that the transformation is not unique. This procedure allows us to shift the integration variable to the rightmost argument of the GPLs, enabling us to perform all the iterated integrals and express the final results in terms of GPLs. We provide the results of $I^{(1)}_{1,1,1}$ in the ancillary file {\tt <3prop1mass.m>}\,.

By applying partial fraction decomposition to the propagators, angular integrals with multiple massive momenta can be reduced to integrals involving a single massive momentum. This implies that integrals involving only massless momenta or a single massive momentum are sufficient to determine all other integrals. The key to the partial fraction technique lies in the fact that angular integrals are linear in parameter space when expressed in a suitable coordinate system. These parameters are typically constructed as ratios of momenta. In the appendix, we provide these relations for three denominator integrals.

\section{Recursion relations}
\label{sec:recursion}
The angular integrals with higher powers of denominators can be related to lower ones through a set of recursion relations, which follow from the equations:
\begin{align}
    \frac{\partial}{\partial v_{kl}} I^{(n)}_{j_1,j_2,j_3} = \sum_{i_1,i_2,i_3} C_{kl}^{(i_1 i_2 i_3)} I^{(n)}_{i_1,i_2,i_3}.
\end{align}
Here, the coefficients $C_{kl}^{(i_1 i_2 i_3)}$ are functions of $v_{ij}, \epsilon, i_1, i_2$ and $i_3$. By utilizing this set of equations, we can express all angular integrals with higher powers of denominators in terms of a finite set of integrals, known as master integrals. These relations are analogous to integration-by-parts identities~\cite{Chetyrkin:1981qh,Laporta:2000dsw}. In ref.~\cite{Lyubovitskij:2021ges}, these relations are explicitly derived for angular integrals with two denominators. For readers' convenience, we provide the complete set of recursion relations, consisting of six linear equations, in the appendix.

\section{Singularity resolution and phase-space integrals}
The angular integrals must be combined with the corresponding radial part to compute the complete phase-space integral. While the functional form of the radial part depends on the specific process, the angular part can be considered process-independent, allowing the result of the angular integration to be applied across various cases. The final phase-space result comes from performing a single parametric integration over the angular part convoluted with the radial part. The integration parameter is typically constructed from dimensionless ratios of the Mandelstam variables.

If the parametric integration contains singularities along the integration path, it is not possible to naively expand the integrand in $\epsilon$ to perform the integration. Therefore, the use of the $\epsilon$-expanded angular integrals requires careful treatment. The singularities from the angular integration are generally collinear in nature and are regularized through the $\epsilon$ parameter. Any additional singularities arising from the parametric integral are typically soft. Consequently, if the $\epsilon$-expanded angular part does not introduce singularities along the integration path, it is safe to use the expanded form order by order.

We observe that for massive angular integrals, the coefficients in the $\epsilon$ expansion do not introduce singularities along the parametric integration path, meaning that the soft singularities arise only from the radial part. However, for massless angular integrals, these parametric singularities can also appear within the individual $\epsilon$-expansion coefficients of the angular part. Since these singularities are soft in nature, we conjecture that they can be resummed to all orders and factored out of the complete $\epsilon$-expansion of the angular part. We have validated this conjecture for the two-propagator case in ref.~\cite{Ahmed:2024x3}.

By correctly factoring out the soft singular part from the angular integral and regulating it using $\epsilon$ through pole subtraction, the full $\epsilon$-expanded expression can be safely used order by order to compute the complete phase-space integrals.

\section{Conclusion}

Phase-space integrals are crucial for achieving precise calculations of observables in particle physics. In this letter, we introduce a novel method for calculating these integrals analytically. First, we present a technique for computing the angular component based on the evaluation of multifold Mellin-Barnes integrals, which commonly arise in various physics scenarios but pose significant challenges for analytic evaluation. For the first time, we provide the analytical results for angular phase-space integrals involving three denominators.

We present results for integrals with all massless momenta up to  $\mathcal{O}(\epsilon^2)$  and for integrals with a single massive momentum up to  $\mathcal{O}(\epsilon)$, expressed in terms of Goncharov polylogarithms. Additionally, we derive a set of recursion relations that reduce integrals with higher powers of denominators to those with lower powers. Utilizing partial fraction decomposition of the propagators, we also introduce a set of linear relations that allows the calculation of angular integrals involving two or more massive momenta. We provide our findings as an ancillary file.

To combine the angular and radial components, a careful treatment of singularities is required, which we detail in this letter. This powerful method not only handles phase-space integrals for massless external particles but also extends to those with massive particles. These integrals are essential for advancing beyond next-to-next-to-leading order (NNLO) accuracy in many scattering processes, such as semi-inclusive deep inelastic scattering.

An independent calculation of the angular part is carried out in ref.~\cite{Huag:2024x3} and we have a perfect numerical agreement. 

\section*{Acknowledgements}
The work of TA and AR is supported by the Deutsche Forschungsgemeinschaft (DFG) through Research Unit FOR2926, \textit{Next Generation Perturbative QCD for Hadron Structure: Preparing for the Electron-Ion Collider}, project number 409651613. Part of SMH’s work was completed during a visit to the University at Buffalo. The authors thank Mainz Institute for Theoretical Physics where a part of the work is carried out. The authors also thank Saurav Goyal, Vaibhav Pathak and Vajravelu Ravindran for valuable discussions and Roman N. Lee, Sven-Olaf Moch, and Narayan Rana for their collaboration on related work~\cite{Ahmed:2024x3}.

\appendix*
\section{Recursion relations}
\label{supp-sec:recursion}
The recursion relations for three-denominator angular  integrals to relate higher powers to lower ones read as
 \begin{align}
 0=&(-D^{\prime}_{5}+D_{6}) I_{-1 + j, k, 1 + l} + D_7 I_{-1 + j, 1 + k, l}\nonumber\\
 &+ (D_5-D^{\prime}_{6}) I_{j, -1 + k, 1 + l} + (D_1 - D^\prime_1) I_{j, k, 
     l}\nonumber\\
     &+ (-D_4 + D^\prime_4) I_{j, k, 1 + l} +(D_{2}-D^{\prime}_{3})I_{1+j,k,l}
    \nonumber\\
  & + (D_3 - D^\prime_2) I_{j, 1 + k, l} - 
   D^\prime_7 I_{1 + j, -1 + k, l},\nonumber\\
%
   0=&E_6 I_{-1 + j, k, 1 + l} + E_7 I_{-1 + j, 1 + k, l}\nonumber\\
   &+ 
  E_5 I_{j, -1 + k, 1 + l} + (E_1 - E^\prime_1) I_{j, k, l} \nonumber\\
  &+ (-E^\prime_2 + E_4) I_{
    j, k, 1 + l}
    - 
  E^\prime_7 I_{j, 1 + k, -1 + l}\nonumber\\
  &+ (E_3 - E^\prime_3) I_{j, 1 + k, l} - 
  E^\prime_5 I_{1 + j, -1 + k, l}\nonumber\\
  &- 
  E^\prime_6 I_{1 + j, k, -1 + l} + (E_2 - E^\prime_4) I_{1 + j, k, l}  ,\nonumber\\
%
 0=&(j + k + l - 1 + 2 \epsilon) I_{j, k, l} - (2 j + l + k + 2 \epsilon) I_{
    1 + j, k, l}\nonumber\\
    &+ v_{11} (j + 1) I_{2 + j, k, l} - l I_{j, k, 1 + l}\nonumber\\
    &- 
  k I_{j, 1 + k, l} + l v_{13} I_{1 + j, k, 1 + l}+ 
  k v_{12} I_{1 + j, 1 + k, l},\nonumber\\
%
 0=&(-1 + 2 \epsilon + j + k + l) I_{j, k, l} - 
  l I_{j, k, 1 + l}\nonumber\\
  &+ (-2 \epsilon - j - 2 k - l) I_{j, 1 + k, l} + 
  l v_{23} I_{j, 1 + k, 1 + l}\nonumber\\
  &+ (1 + k) v_{22} I_{j, 2 + k, l} - 
  j I_{1 + j, k, l}\nonumber\\
  &+ j v_{12} I_{1 + j, 1 + k, l}    ,\nonumber\\
%
  0=&(-1 + 2 \epsilon + j + k + l) I_{j, k, l}\nonumber\\& + (-2 \epsilon - j - k - 2 l) I_{j, 
    k, 1 + l}\nonumber\\
    &+ (1 + l) v_{33} I_{j, k, 2 + l} - k I_{j, 1 + k, l}\nonumber\\
    &+ 
  k v_{23} I_{j, 1 + k, 1 + l} - j I_{1 + j, k, l} \nonumber\\
  &+ 
  j v_{13} I_{1 + j, k, 1 + l}   ,\nonumber\\
%
 0=&(-1 + 2 \epsilon + j + k + l) I_{j, k, l} - l I_{j, k, 1 + l} \nonumber\\
 &- 
  k I_{j, 1 + k, l} + (-2 \epsilon - 2 j - k - l) I_{1 + j, k, l}\nonumber\\
  &+ 
  l u_{13} I_{1 + j, k, 1 + l} + 
  k u_{12} I_{1 + j, 1 + k, l}\nonumber\\
  &+ (1 + j) u_{11} I_{2 + j, k, l}.
 \end{align}
The coefficients $\{D_i, D'_i, E_i, E'_i\}$ are rational functions of the kinematic variables $v_{kl}$, with their explicit forms detailed in the ancillary file {\tt <{recursion.m}>}\,.

\section{Three denominator integral with double and triple massive momenta} 
We present the relations that express any double-(triple-)massive, three-denominator phase space angular integral with arbitrary denominator powers $j, k, l$ as a linear combination of single-massive, three-denominator phase space angular integrals. The double-massive integral is decomposed as
%
\begin{align}
&I^{(2)}_{j,k,l}(v_{11},v_{22},v_{12},v_{13},v_{23})=\nonumber\\
&\sum_{n=0}^{j-1}
\begin{pmatrix}
k-1+n\\
k-1
\end{pmatrix}
\lambda^{k}_{\pm}(1-\lambda_{\pm})^{n}I^{(1)}_{j-n,k+n,l}(v_{11},v^{\pm}_{13},v_{13},v^{\pm}_{33})\nonumber\\
&+\sum_{n=0}^{k-1}
\begin{pmatrix}
j-1+n\\
j-1
\end{pmatrix}
\lambda_{\pm}^{n}(1-\lambda_{\pm})^{j}I^{(1)}_{k-n,j+n,l}(v_{22},v^{\pm}_{23},v_{23},v^{\pm}_{33}).
\end{align}
The parameters $\lambda_{\pm}, v^{\pm}_{13}, v^{\pm}_{33}, v^{\pm}_{23}$ are functions of the original variables $v_{11}, v_{22}, v_{12}, v_{13}, v_{23}$ from the double-massive, three-denominator integral, and are given by
\begin{align}
\lambda_{ \pm}=&\frac{v_{12}-v_{11} \pm \sqrt{v_{12}^2-v_{11} v_{22}}}{2 v_{12}-v_{11}-v_{22}},\nonumber\\
v_{13}^{ \pm}=&\left(1-\lambda_{ \pm}\right) v_{11}+\lambda_{ \pm} v_{12},\nonumber\\
v_{23}^{ \pm}=&\left(1-\lambda_{ \pm}\right) v_{12}+\lambda_{ \pm} v_{22},\nonumber\\
v^{\pm}_{33}=&(1-\lambda_{\pm})v_{13}+\lambda_{\pm}v_{23}.
\end{align}
%
%

The integral with three massive momenta can be derived from those involving a single massive momentum through
\begin{align}
&I^{(3)}_{j,k,l}(v_{11},v_{22},v_{33},v_{12},v_{13},v_{23})=\nonumber\\
&\sum_{n=0}^{j-1}
\begin{pmatrix}
l-1+n\\
l-1
\end{pmatrix}
\lambda^{l}_{\pm}(1-\lambda_{\pm})^{n}\Bigg\{\sum_{m=0}^{j-n-1}
\begin{pmatrix}
k-1+m\\
k-1
\end{pmatrix}
\nonumber\\
&\times\tilde{\lambda}_{\pm}^{k}(1-\tilde{\lambda}_{\pm})^{m}I^{(1)}_{j-n-m,l+n,k+m}(v_{11},v_{13}^{\pm},\tilde{v}_{13}^{\pm},\tilde{v}^{\prime\pm}_{33})\nonumber\\
&+\sum_{m=0}^{k-1}
\begin{pmatrix}
j-n-1+m\\
j-n-1
\end{pmatrix}
\tilde{\lambda}^{m}_{\pm}(1-\tilde{\lambda}_{\pm})^{j-n}\nonumber\\
&\times I^{(1)}_{k-m,l+n,j-n+m}(v_{33},v_{33}^{\pm},\tilde{v}_{33}^{\pm},\tilde{v}^{\prime\pm}_{33})
\Bigg\}\nonumber\\
&+\sum_{m=0}^{l-1}
\begin{pmatrix}
j-1+n\\
j-1
\end{pmatrix}
\lambda^{n}_{\pm}(1-\lambda_{\pm})^{j}\Bigg\{\sum_{m=0}^{l-n-1}
\begin{pmatrix}
k-1+m\\
k-1
\end{pmatrix}
\nonumber\\
&\times\lambda^{\prime k}_{\pm}(1-\lambda^{\prime}_{\pm})^{m}I^{(1)}_{l-n+m,j+n,k+m}(v_{22},v_{23}^{\pm},v_{23}^{\prime \pm},v^{\prime\prime\pm}_{33})\nonumber\\
&+\sum_{m=0}^{k-1}
\begin{pmatrix}
l-n-1+m\\
l-n-1
\end{pmatrix}
\lambda^{\prime m}_{\pm}(1-\lambda^{\prime}_{\pm})^{l-n}\nonumber\\
&\times I^{(1)}_{k-m,j+n,l-n+m}(v_{33},v_{33}^{\pm},v_{33}^{\prime\pm},v^{\prime\prime\pm}_{33})
\Bigg\},
\end{align}
where 
\begin{align}
\lambda_{ \pm}=&\frac{v_{12}-v_{11} \pm \sqrt{v_{12}^2-v_{11} v_{22}}}{2 v_{12}-v_{11}-v_{22}}, \nonumber\\
\tilde{\lambda}_{ \pm}=&\frac{v_{13}-v_{11} \pm \sqrt{v_{13}^2-v_{11} v_{33}}}{2 v_{13}-v_{11}-v_{33}}, \nonumber\\
\lambda^{\prime}_{ \pm}=&\frac{v_{23}-v_{22} \pm \sqrt{v_{23}^2-v_{22} v_{33}}}{2 v_{23}-v_{22}-v_{33}}, \nonumber\\
v_{13}^{ \pm}=&\left(1-\lambda_{ \pm}\right) v_{11}+\lambda_{ \pm} v_{12},\nonumber\\
 v_{23}^{ \pm}=&\left(1-\lambda_{ \pm}\right) v_{12}+\lambda_{ \pm} v_{22},\nonumber\\
v^{\pm}_{33}=&v_{13}(1-\lambda_{\pm})+\lambda_{\pm}v_{23},\nonumber\\
%
 \tilde{v}^{\pm}_{13}=&\frac{1}{v_{11}+v_{33}-2 v_{13}}\Bigg(v_{11}\left(v_{33} \pm \sqrt{v_{13}^2-v_{11} v_{33}}\right)\nonumber\\
&-v_{13}\left(v_{13} \pm \sqrt{v_{13}^2-v_{11} v_{33}}\right)\Bigg), \nonumber\\
\tilde{v}^{\pm}_{33}=&\frac{1}{v_{11}+v_{33}-2 v_{13}}\Bigg(v_{33}\left(v_{11} \mp \sqrt{v_{13}^2-v_{11} v_{33}}\right)\nonumber\\
 &-v_{13}\left(v_{13} \mp \sqrt{v_{13}^2-v_{11} v_{33}}\right)\Bigg),\nonumber\\
 v^{\prime \pm}_{23}=&\frac{1}{v_{22}+v_{33}-2 v_{23}}\Bigg(v_{22}\left(v_{33} \pm \sqrt{v_{23}^2-v_{22} v_{33}}\right)\nonumber\\
&-v_{23}\left(v_{23} \pm \sqrt{v_{23}^2-v_{22} v_{33}}\right)\Bigg),\nonumber\\
v^{\prime \pm}_{33}=&\frac{1}{v_{22}+v_{33}-2 v_{23}}\Bigg(v_{33}\left(v_{22} \mp \sqrt{v_{23}^2-v_{22} v_{33}}\right)\nonumber\\
 &-v_{23}\left(v_{23} \mp \sqrt{v_{23}^2-v_{22} v_{33}}\right)\Bigg),\nonumber\\
\tilde{v}^{\prime \pm}_{33}=&v_{11}(1-\lambda_{\pm})(1-\tilde{\lambda}_{\pm})+v_{13}\tilde{\lambda}_{\pm}(1-\lambda_{\pm})\nonumber\\
&+v_{12}\lambda_{\pm}(1-\tilde{\lambda}_{\pm})+\lambda_{\pm}\tilde{\lambda}_{\pm}v_{23},\nonumber\\
v^{\prime\prime \pm}_{33}=&v_{12}(1-\lambda_{\pm})(1-\lambda^{\prime}_{\pm})+v_{13}\lambda^{\prime}_{\pm}(1-\lambda_{\pm})\nonumber\\
&+v_{22}\lambda_{\pm}(1-\lambda^{\prime}_{\pm})+v_{23}\lambda_{\pm}\lambda^{\prime}_{\pm}.
\end{align}


\bibliography{main}

\end{document}